\documentclass[sigconf]{acmart}

\AtBeginDocument{%
  \providecommand\BibTeX{{%
    \normalfont B\kern-0.5em{\scshape i\kern-0.25em b}\kern-0.8em\TeX}}}

\usepackage{graphicx}
\usepackage{caption}
\usepackage{totcount}
\regtotcounter{page}

\copyrightyear{2024}
\acmYear{2024}
\setcopyright{rightsretained}
\acmConference[UIST Adjunct '24]{The 37th Annual ACM Symposium on User Interface Software and Technology}{October 13--16, 2024}{Pittsburgh, PA, USA}
\acmBooktitle{The 37th Annual ACM Symposium on User Interface Software and Technology (UIST Adjunct '24), October 13--16, 2024, Pittsburgh, PA, USA}
\acmDOI{10.1145/3672539.3686326}
\acmISBN{979-8-4007-0718-6/24/10}

\begin{document}

\title{Improving Interface Design in Interactive Task Learning for Hierarchical Tasks based on a Qualitative Study}

\author{Jieyu Zhou}
\affiliation{%
  \institution{Georgia Institute of Technology}
  \city{Atlanta}
  \state{Georgia}
  \country{USA}
}
\email{jzhou625@gatech.edu}

\author{Christopher MacLellan}
\affiliation{%
  \institution{Georgia Institute of Technology}
  \city{Atlanta}
  \state{Georgia}
  \country{USA}
}
\email{cmaclell@gatech.edu}

\begin{abstract}
Interactive Task Learning (ITL) systems acquire task knowledge from human instructions in natural language interaction. The interaction design of ITL agents for hierarchical tasks stays uncharted. This paper studied Verbal Apprentice Learner(VAL) for gaming, as an ITL example, and qualitatively analyzed the user study data to provide design insights on dialogue language types, task instruction strategies, and error handling. We then proposed an interface design: Editable Hierarchy Knowledge (EHK), as a generic probe for ITL systems for hierarchical tasks.
\end{abstract}

\keywords{Hierarchical Task, Task Agent}

\maketitle

\section{Introduction}

Task-oriented agents are widely used across various domains \cite{wang2024survey,laird2017interactive,maclellan2018framework}, including smart speakers (Alexa, Google Home, Siri), multi-model devices (Rabbit R1 \cite{rabbit2023}, AI Pin), web-based systems \cite{huq2023s,li2023using, li2017sugilite}, and game controls \cite{wang2023voyager,wang2023describe,lawley2023val}. They communicate between different environments, by translating user requests to application programming interfaces (API) for task execution. User's natural language input is often syntactically loose and ambiguous, but the APIs require rigid, specific actionable commands in the programming language. Unlike chatbots, which simply answer the query in an isolated environment, with limited ability to address the challenges of translating loose natural language to rigid API commands, task-oriented agents guarantee the robustness of the translation and the correctness of execution.

Moreover, when the built-in API commands are insufficient for a user's instruction, task-oriented agents should learn from the user to customize the API commands. Interactive Task Learning (ITL) system is one approach to this issue. ITL agents acquire task knowledge, such as task names and their subtasks, from human instructions provided via natural language interaction \cite{laird2017interactive, gluck2019interactive}. Interaction with ITL systems for relatively simple tasks, such as "ordering Cappuccino on Starbucks website", has been widely explored in HCI field. For example, ONYX conducts data visualization works \cite{ruoff2023onyx}, SOVITE refers to third-party apps to repair conversational breakdowns implementing on flight booking \cite{li2020multi}, and Sugilite enables users to create automation on Android systems \cite{li2017sugilite}. However, when ITL systems face complicated tasks with multiple steps and hierarchical structures \cite{mininger2022demonstration, kirk2024improving, lawley2023val},  which kinds of interaction are suitable needs to be further explored.

Verbal Apprentice Learner (VAL) \cite{lawley2023val} is one prior work that explored user interaction for complicated tasks. VAL is a neuro-symbolic ITL system, capable of interactively acquiring task knowledge from natural language instruction provided by end users tested in game environments \cite{lawley2023val}.  It is neuro by leveraging LLMs to address the challenges associated with parsing error-prone natural language inputs, a common issue in prior ITL systems. On the other hand, it is symbolic by engaging in multi-step reasoning through a hierarchical task network (HTN) planning system \cite{baier2009htn} and grounding a task to default primitive actions (executable API commands). The original VAL study \cite{lawley2023val} focused primarily on evaluating the accuracy and performance of VAL's components. However, the interaction transcripts and open-ended survey results have never been analyzed. In this paper, we analyzed the data, generating design insights and new VAL 2.0 prototypes. We believe this work shall also provide insights for other ITL systems for complicated hierarchical tasks.

\section{Design Insights}
We analyzed the log data of the user study \cite{lawley2023val}, where 12 participants were asked to teach VAL to make onion soup in the Overcooked game. After the study, there was a post-session survey, consisting of seven-point Likert-scale questions and open-ended questions asking users comments based on their experience. We applied a reflexive thematic analysis approach \cite{braun2006using} on 1336 pieces of log data paired with participants' survey results.

\subsection{Use the type of language desired from users}
We found participants used two distinct types of language (some used both): 1. natural language (12 participants used), similar to human communication in daily life. Example: "Could you please go grab an onion and move to the onion storage tile? Do you know where onion storage tile is?" (P1); 2. coding language (5 participants used), has the format "action(object)", such as make(onion, pot), pressSpace(onion). VAL mostly interacts in natural language but also uses coding language, such as "I have learned the action: cook(onion)" in some scenarios. This sometimes leads participants to mirror coding language from VAL. Unfortunately, VAL cannot parse coding language very well, which leads to errors.

In general, task-oriented agents act as a connection between users' inputs in natural language and concise API command output in coding language. The designer shall avoid directly displaying the back-end API commands in the front-end dialog, as users mirror the language used by the agent. Therefore, to avoid mode errors \cite{norman2013design}, the dialog should keep only using the natural language type.

\subsection{Handle different task instruction strategies}
Users may apply different strategies to teach the ITL agent to conduct a complicated task hierarchically. A typical task hierarchy of cooking onion soup is shown on the right side of Figure 1. In VAL, We identified three strategies to form the hierarchy: 1. top-down (8 participants used), beginning with the task name and listing all the sub-steps; 2. bottom-up (3 participants used) starting with detailed steps and then combining into a higher-level task, and 3. intermediate (1 participant used) starting with middle-level tasks of cooking onion soup, such as "boiling the onion", "plating the soup" and then grounding them out to primitive actions (press space and move to).

We observed that bottom-up strategies sometimes cause a shallow hierarchy, which only contains one layer (the main task and a sequence of primitive actions under it), especially when the user describes the task steps in a very concise way without any redundant words, such as "Move to onion, press space, move to pot, press space, press space" (P3). We perceive a shallow hierarchy as a learning failure because it compromises the generalizability of the learning task. If a user wants to teach VAL a new task containing subtasks it already knows, a deep hierarchy can reuse some of its intermediate steps, while a flat hierarchy must limits reusability. The design insight drawn from this issue is to ask the users to clarify the higher-level tasks when a flat hierarchy is detected.

\subsection{Offer multiple ways for error handling}
Error handling is a major requirement for task-oriented agents, they should prevent an error before it happens or recover from it after it occurs. Unlike single-or-few-step tasks, whose error handling has been discussed \cite{li2017sugilite,li2020multi}, the complicated multi-step scenario remains uncharted. In multi-step tasks, an error in one step can propagate and eventually compromise executing the whole task. Besides execution errors faced by all other task-oriented agents, ITL agents may encounter extra potential errors in parsing due to its interactive nature (i.e., mistakes in the task names or objects). 

One approach is to prevent errors by confirming the user instruction in a dialog. However, frequent confirmation dialog can frustrate the user. This is particularly true for multi-step task ITLs where the dialog can be overwhelming for the user because each parsing contains multiple input confirmations and each task has multiple steps. From VAL's log data, participants experienced an average of 45 confirmations to make onion soup, and they complained: "\textit{Although I found it helpful to check that VAL parsed my instructions correctly, it also felt tedious to answer them all the time.}"(P11). How might we reduce the tedious confirmation without risking the system's correctness? One approach is to compensate by providing easy recoveries after any error occurs. Balancing error prevention and recovery remains critical to designing a robust ITL system.

VAL has a simple recovery strategy: an option to undo instructions which was used by 3/4 participants in the tests. However, a straightforward undo is not enough, because users do not often notice their mistakes immediately and continue making commands. After noticing an error, they must also undo the sequential correct instructions. As one participant said: "\textit{It was difficult to undo or go back. I first told VAL an incorrect sequence, and it was not easy to re-teach VAL from scratch.}" (P12). Designers of ITL systems for multi-step tasks should offer multiple error-handling strategies that account for the interrelationship of steps and the balance between prevention and recovery to avoid tediousness.

\section{Design Prototypes}
Based on the design insights, we propose VAL 2.0 which fetures a revised the dialog interface and a new Graphic User Interface (GUI), called Editable Hierarchy Knowledge (EHK), targeted specifically for complicated hierarchy tasks, as shown in Figure 1. 

\begin{figure}[h]
  \centering
  \includegraphics[width=\linewidth]{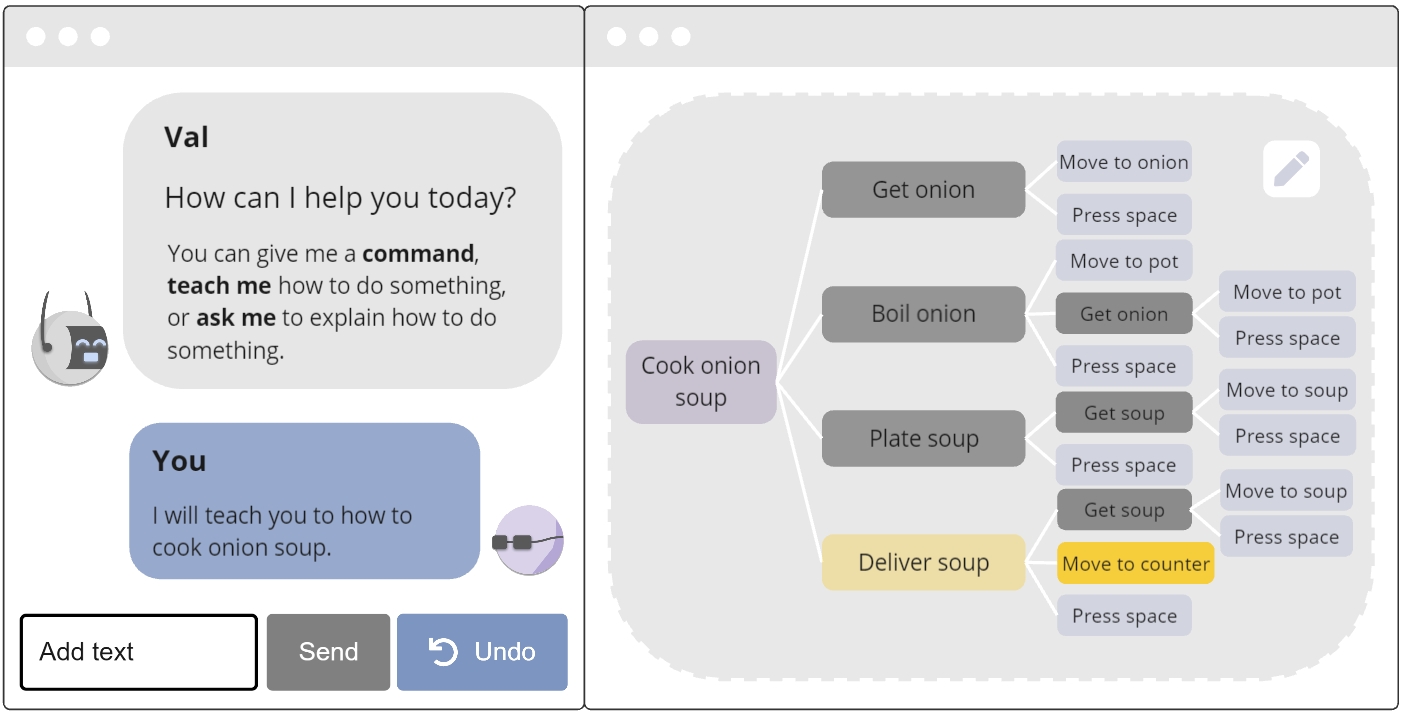}
  \caption{Example of Editable Hierarchy Knowledge GUI} 
  \Description[Figure 1 shows an example of Editable Hierarchy Knowledge GUI]{The left side of the figure shows the dialog that a user teaching VAL to cook onion soup, and the right side presents the hierarchy of that task, which can be editable}
\end{figure}

First, we unified the language type. Particularly, we now describe the primitive tasks and known actions in natural language, reducing user's use of coding language that VAL cannot parse.

Second, we want VAL 2.0 to better fit different task instruction strategies. The proposed EHK is specially designed to extract and present the hierarchical structure from the user's instructions. It highlights the current step and folds previous subtasks in synchronous with the teaching process. The EHK interface provides granular feedback to the users that they are understood by the agent \cite{ruoff2023onyx, amershi2019guidelines}. If a flat hierarchy is detected, VAL will prompt a request for higher-level tasks, with EHK background flash as a visual alert.

Third, ongoing efforts are made to VAL 2.0 for better error handling by balancing between error prevention and recovery. To reduce the cumbersome confirmation dialogues, we added features to skip certain types of dialog. Instead, we offer several error recovery methods. VAL 2.0 allows users to revise the instruction directly via natural language in addition to undo. Moreover, in the EHK interface, users can delete an action or drag it to change the hierarchical structure of the task. We believe the EHK interface can be a way to compensate for the cost of reducing confirmation dialog frequency.

In the future, more tests will be done to test the usability of these new features, especially for the EHK interface. We believe the design insights and how we apply them to iterative design can inspire other ITL systems with complicated hierarchical tasks.

\bibliographystyle{ACM-Reference-Format}
\bibliography{source}

\end{document}